# A WALL-FUNCTION APPROACH FOR DIRECT PRECIPITATION/CRYSTALLIZATION FOULING IN CFD MODELLING

**S.G. Johnsen, S.T. Johansen and B. Wittgens**

SINTEF Materials and Chemistry, Trondheim, Norway,
E-mail (Corresponding author): sverre.g.johnsen@sintef.no

**ABSTRACT**

The main objective of this paper is to present a generic modelling framework, for the diffusive mass transport through the turbulent, reactive boundary layer of multi-component fluid mixtures that precipitate on the wall. The modelling is based on Maxwell-Stefan diffusion in multi-component mixtures, the relaxation to chemical equilibrium model, and the single-phase Navier-Stokes equations. Finally, turbulence is introduced by Reynolds-averaging. The governing equations are simplified in accordance with common assumptions of computational fluid dynamics (CFD), and based on the assumption that the over-all bulk flow is parallel to the wall, 1-dimensional equations for the species and heat transport perpendicular to the wall have been formulated. The equations are solved on a fine grid in order to fully resolve the boundary layer, and the effect of allowing/disallowing deposition and chemical reactions was investigated for a simplified test-case (4-component ideal mixture of perfect gasses capable of chemical reaction and sublimation fouling).

The developed framework can be employed as a sub-grid model for direct precipitation/crystallization/solidification fouling in coarse grid CFD models.

**INTRODUCTION**

Fouling of solid surfaces and heat-exchanger surfaces in particular, is a common and much studied problem in most process industries, as reflected in the recent review paper by Hans Müller-Steinhagen (2011). Fouling is defined as the unwanted accumulation of solid (or semi-solid) material on solid surfaces. Similar is the desired accumulation of solids e.g. in chemical vapor deposition (Kleijn et al., 1989; Krishnan et al., 1994).

A common and costly problem in many industrial applications is the direct precipitation of super saturated fluids on heat-exchanger surfaces. Typical examples are found in e.g. the high-temperature off-gas from waste incineration, metal production, or in power plants, where efficient heat recovery is key to sustainable production, and where a combination of direct precipitation and deposition of e.g. solid metal oxides is a major show-stopper.

By precipitation we understand all types of phase transitions from a fluid to a relatively denser phase, e.g. gas $\rightarrow$ liquid (condensation), gas $\rightarrow$ solid (sublimation), liquid $\rightarrow$ solid (solidification). For some materials, the precipitate may have a crystalline structure (crystallization).

In our modelling work, fouling due to mass deposition from a fluid phase is grouped into two different classes; 1) particulate fouling, where particles carried by the fluid phase penetrate the laminar boundary layer and stick to the wall (e.g. precipitates, dust, or soot particles) (previous publications); and 2) direct precipitation where the fluid becomes super-saturated close to the wall and a phase-transition occurs at the wall (current paper). The direct precipitation on solid surfaces is due to the molecular diffusion through the stagnant boundary layer close to the wall. This is a complex physical process where the diffusion flux of each species is coupled to the diffusion fluxes and thermodynamic/chemical properties of all the species present. Thermal gradients and potential for chemical reactions or phase-change add to the complexity. In the view of thermodynamics, direct precipitation may occur when the phase transition is energetically favorable in terms of minimizing the total Gibbs free energy of the fluid-precipitate system. Commonly, a combination of 1 and 2 takes place. Fouling can only occur if the adhesive forces between the foulant and the wall are strong enough to overcome the flow-induced shear forces at the wall.

In previous papers (Johansen, 1991; Johnsen and Johansen, 2009a; b; Johnsen et al., 2010), we developed a framework for the mathematical modelling of particle deposition and re-entrainment and demonstrated how this model could be employed as a wall boundary condition (mass sink) for CFD models.

In the current work, we present a generic framework for the mathematical modelling of the molecular transport through the turbulent, reactive boundary layer of multi-component fluid mixtures that precipitate on the wall. The modelling of chemically reacting boundary-layer flow has been discussed in detail by e.g. R. A. Baurle (2004) and more recently by Kuo and Acharya (2012). Direct precipitation is characterized by the molecular transport and deposition of one or more of the species in the fluid phase. Thus, models for the transport through the turbulent boundary layer must be established for each species in the fluid. The current modelling work starts out with the general single-phase Navier-Stokes equations, introduces turbulence and dimensionless variables/parameters, and makes simplifications/

Table 1. Governing equations

| | |
|---|---|
| $\overline{\nabla}\left(\rho_f X_i \vec{u}_f\right) + \overline{\nabla}\vec{j}_{d,i} = R_i$ | (1) |
| $\overline{\nabla}\left(\rho_f \vec{u}_f \vec{u}_f\right) = -\overline{\nabla}P + \overline{\nabla}\boldsymbol{\tau} + \rho_f \vec{g}$ | (2) |
| $\overline{\nabla}\left(\rho_f h_{sens} \vec{u}_f\right) = k_f \overline{\nabla}T - \sum_i \vec{j}_{i,d} h_{i,sens} - \sum_i \left(\Delta h_f\right)_i^0 R_i$ | (3) |
| $\sum_i X_i = \sum_i z_i = 1$ | (4) |

assumptions appropriate in the boundary layer. The resulting framework is built on differential equations for the mass-fraction of each species, and the velocity and temperature of the fluid, which are discretized and solved numerically on a 1-dimensional mesh. The formulation is designed for implementation as a sub-grid wall-function for coarse-grid CFD-models, and can give significant reduction in computational costs compared to modelling the boundary-layer directly with a fine-grid CFD model.

Due to its generic nature, the modelling framework requires appropriate sub-models for the underlying physics; e.g. chemical reaction and thermodynamic kinetics, transport parameters, and turbulence. As a test-case, we demonstrate the applicability of the modeling framework to a simplified 4-component ideal gas mixture, where three of the components participate in a reversible chemical reaction, whereas the fourth component is inert. The third component, which is the product of the chemical reaction, sublimates and deposits on the wall. For this test-case, we show the impact of turning on/off deposition and chemical reactions.

## MATHEMATICAL MODELLING FRAMEWORK

We consider a multi-component fluid mixture, consisting of $N$ chemical species, advected at the local mass-averaged velocity $\vec{u}_f$, close to a wall.

For simplicity, we consider single-phase flow only, such that there are no phase transitions except at the wall, where precipitation may take place.

### Governing Equations

The set of steady-state governing equations consists of the Advection-Diffusion equation for each species, the fluid mixture momentum and energy equations, and the restriction that the mass- and mole-fractions must sum to unity (see Table 1).

### Diffusive Mass Flux Density

Molecular diffusion is modelled by the Maxwell-Stefan method (see e.g. (Taylor and Krishna, 1993; Krishna and Wesselingh, 1997)). In an isobaric situation where the chemical potential depends on temperature and mixture composition only, the diffusive mass flux density of the species $i \in \{1,\ldots,N-1\}$ is expressed as

$$\vec{j}_{i,d} = -\rho_f D_{ij}\left[d_{T,j}\overline{\nabla}\ln T + \Gamma_{jk}\Lambda_{kl}\overline{\nabla}X_l\right], \quad (5)$$

where the elements of the inverse diffusivity tensor are given by $D^{-1}_{ii} = M_{w,f}/\left(Ð_{iN}M_{w,N}\right) + \sum_{l \neq i}\left(z_l M_{w,f}\right)/\left(z_i Ð_{il} M_{w,i}\right)$ and $D^{-1}_{i,j \neq i} = -M_{w,f}/\left(Ð_{ij}M_{w,j}\right)$, the thermophoretic driving force coefficients are given by $d_{T,j} \equiv \partial_T G_{z,f,j}/\mathcal{R} = -\partial_{z_j}S_f/\mathcal{R}$, the elements of the diffusiophoretic driving force coefficient tensor are given by $\Gamma_{jk} = \partial_{z_k}G_{z,j}/\mathcal{R}T = \delta_{jk}/z_j + \partial_{z_k}\ln\gamma_j$, the elements of the mole-mass-fraction conversion tensor are given by $\Lambda_{kl} = \left(M_{w,f}/M_{w,N}\right)z_k + \left(M_{w,f}/M_{w,l}\right)(\delta_{kl} - z_k)$, and Einstein's summation convention is employed.

### Transport parameters

Due to the varying composition and temperature through the turbulent boundary layer, a fluid mixture with variable flow parameters must be considered. Thus, sub-models are required for the mass-density, activity coefficients, viscosity, thermal conductivity, specific heat capacity, binary diffusivities, and entropy. The mass density and activity coefficients are determined by the thermodynamics of the fluid mixture and are generally functions of temperature, pressure and fluid composition. They must be estimated from an equation of state or from experimental data. The mixture viscosity, thermal conductivity, heat capacity and entropy can be estimated from e.g. correlations or experimental data (see example test-case in the appendix).

### Chemical Reaction Kinetics

Due to the difficulties in performing a standard averaging procedure, several modelling concepts have been developed for turbulent reacting flows. We employ the *relaxation to chemical equilibrium model* (Myhrvold, 2003; ANSYS Customer Portal, 2015). This simplified model is a good approximation in the case of mixing-limited chemical reaction, i.e. when the time-scale of mixing is much longer than the time-scale of chemical reaction. The net reaction rate of species $i$ only depends on the deviation from chemical equilibrium for that species, alone;

$$R_i = \rho_f\left(X_i^{eq} - X_i\right)/\tau_{char}, \quad (6)$$

where $X_i^{eq}$ is the mass fraction of the $i$th species at chemical equilibrium, and $\tau_{char}$ is a characteristic time-scale, which in the mixing-limited regime is equal to the characteristic time-scale of the diffusion/turbulent dispersion.

### Turbulence

Turbulence is introduced by expressing the flow-variables as the sum of an ensemble average and a deviatoric term that averages to zero, and ensemble (Reynolds) averaging the governing equations. An arbitrary flow variable can thus be expressed as $\phi = \bar{\phi} + \phi'$. The fluid velocity is expressed in terms of the mass-weighted Favre average, $\vec{u}_f = \widetilde{\vec{u}}_f + \vec{u}''_f$, where $\widetilde{\vec{u}}_f \equiv <\rho_f \vec{u}_f>/<\rho_f>$, the $<\ >$ denotes ensemble averages, and $<\vec{u}''_f> = (v_t/Sc_t)\overline{\nabla}\left(\ln\langle\rho_f\rangle\right)$. The correlations between the deviatoric terms of an advected variable and the fluid velocity, or another advected variable, are modelled by gradient transport approximations, $<\phi' \mathbf{u}''_f> = -(\nu_t/Sc_t)\overline{\nabla}\overline{\phi}$, and $\langle\phi'\psi'\rangle = -(\tau_L \nu_t/Sc_t)(\overline{\nabla}\overline{\phi})(\overline{\nabla}\overline{\psi})$, respectively. The correlation between the deviatoric parts of the sensible enthalpy and the velocity is modelled as $<\mathbf{u}''_f h'_{sens}> = -(\nu_t/Pr_t)\overline{\nabla}\overline{h_{sens}}$. Closure laws are now required for the turbulent kinematic viscosity and turbulent Schmidt and Prandtl numbers. The turbulent kinematic viscosity is modelled as

$$\nu_t^+ \equiv \frac{\nu_t}{\nu_w} = \begin{cases} (y^+/11.15)^3 & \text{for } y^+ < 3.0 \\ (y^+/11.4)^2 - 0.049774 & \text{for } 3.0 \leq y^+ \leq 52.108 \\ 0.4 y^+ & \text{for } 52.108 < y^+ \end{cases}$$

(7)

whereas the turbulent Schmidt and Prandtl numbers and the Lagrangian time-scale were assumed constant, $Sc_t = Pr_t = 1$, and $\tau_L = 11$. For more details, refer to e.g. Johansen (1991). In the following, the averaging notation is suppressed, and all variables/parameters are considered ensemble averaged over turbulent realizations.

**Near-Wall Model**

The governing equations, (1), (2), and (3), are solved in the near-wall region, where we may impose several simplifying assumptions. Furthermore, the aim of this work is to prepare a modelling framework that will later be implemented as a sub-grid model acting as the wall boundary condition (mass sink) for CFD models. This allows for further simplification.

We define the Cartesian y-direction to be normal to the wall, and the unit vector $\hat{\mathbf{y}}$ points into the fluid. The mean bulk flow is in the axial direction. We assume fully developed flow, and all derivatives in the axial and transversal directions are negligible. Furthermore, we assume that all gradients are zero in the bulk.

It is assumed that pressure variations are due to hydrostatic gradients only, that the wall-normal advective velocity is negligible, and that the sensible enthalpy can be approximated by $h_{sens} = c_P T$.

**Dimensionless Variables/Parameters**

In the governing equations, all variables and parameters are made dimensionless by scaling with the value at the wall. Exceptions are the dimensionless velocity $u_f^+ \equiv u_f/u_\tau$, wall distance $y^+ \equiv y u_\tau/\nu_{f,w}$, time-scale $\tau^+ = u_\tau^2 \tau/\nu_{f,w}$, and temperature $T^+ = \frac{(T - T_w) u_\tau \rho_{f,w} C_{P,w}}{q_w}$ where $q_w = -k_f \partial_y T\big|_w$ is the heat flux density at the wall. It is also convenient to introduce the dimensionless groups $T_w^{0+} \equiv T^{wall} u_\tau \rho_{f,w} c_{P,w}/q_w$, $Pr_w \equiv \mu_w c_{P,w}/k_{f,w}$, $Re_{BL} \equiv \rho_w u_\tau y^{bulk}/\mu_w$, and $Nu_w \equiv h y^{bulk}/k_w$.

**Governing Equations Revisited**

Introducing turbulence, dimensionless variables and the simplifications mentioned above, the governing equations (1)-(3) become

$$\partial_{y^+}\left[\frac{\nu_t^+}{Sc_t}\rho_f^+ \partial_{y^+} X_i\right] + \partial_{y^+} j_{d,i,y}^+ + \frac{\rho_f^+}{\tau_{char}^+}\left(X_i^{eq} - X_i\right)$$
$$-\frac{\rho_f^+}{\tau_{char}^+}\left[\frac{\tau_L^+ \nu_t^+}{Sc_t}\left(\partial_{y^+} \ln \rho_f^+\right)\partial_{y^+}\left(X_i^{eq} - X_i\right)\right] = 0 \quad ;$$

(8)

$$\partial_{y^+} u_{f,x}^+ = 1/(\mu^+ + \mu_t^+) \quad ;$$ (9)

$$\frac{1}{Pr_w}\partial_{y^+}\{K_{(0)}^+ T^+\} + \frac{1}{Pr_w}\partial_{y^+}\{K_{(1)}^+ \partial_{y^+} T^+\} + \Theta = 0 \quad ;$$ (10)

where $K_{(0)}^+ \equiv (k_{f,c}^+ + k_{f,t}^+)(\partial_{y^+} \ln c_P^+) - Pr_w \sum_{i=1}^{N} \vec{j}_{i,d,y}^+ c_{P,i}^+$

$K_{(1)}^+ \equiv k_f^+ + k_{f,t}^+ + k_{f,c}^+$, and

$$\Theta \equiv \left(\frac{Pr_w Re_{BL}}{Nu_w}\right)\left(\frac{T_w}{T_b - T_w}\right)\partial_{y^+}\sum_{i=1}^{N}\vec{j}_{i,d,y}^+ c_{P,i}^+$$
$$+\frac{\rho_f^+}{\tau_{char}^+}\sum_{i=1}^{N}(\Delta h_f^+)_i^0\left[(X_i^{eq} - X_i) - \frac{\tau_L^+ \nu_t^+}{Sc_t}(\partial_{y^+} \ln \rho_f^+)\partial_{y^+}(X_i^{eq} - X_i)\right]$$, for

species mass-fractions, dimensionless axial fluid mixture velocity, and dimensionless temperature, respectively. For simplicity, the first term in $\Theta$ was neglected.

**Numerical Model**

The dimensionless governing equations were discretized and solved on a 1-dimensional grid extending from the wall ($y = 0$) to the bulk ($y = y^{bulk}$). The axial velocity is solved by an Euler forward-stepping method, and the coupled mass-fraction and temperature equations are solved simultaneously by an implicit method.

The system of equations resulting from Eqs. (8)-(10) is solved iteratively by the algorithm suggested by Patankar (1980):

1. Initialize $u_{f,x}^+$, $X_i$, $T^+$, $u_\tau$, and $q_w$.
2. Solve for $u_{f,x}^+$, $X_i$ and $T^+$.
3. Update $u_\tau$, $q_w$, fluid properties, and rhs-vectors.
4. Check if the correction of the calculated mass-fraction profile is larger than some convergence criteria. If not converged, return to step 2.
5. The solution is converged.

**Boundary Conditions**

The following Dirichlet boundary conditions (BC) apply to the governing equations:

$$u_{x,w}^+ = 0 \quad ,$$ (11)

$$u_x^+(y^+ = y_b^+) = u_{x,b}^+ \quad ,$$ (12)

$$T_w^+ = 0 \quad ,$$ (13)

$$T^+(y^+ = y_b^+) = T_b^+ \quad ,$$ (14)

$$X_i(y^+ = y_b^+) = X_i^{bulk} \quad .$$ (15)

Table 2. Test-case flow conditions

| | | |
|---|---|---|
| $T^{bulk}$ | 1300 | $K$ |
| $T^{wall}$ | 1200 | $K$ |
| $u_x^{bulk}$ | 10 | $m/s$ |
| $u_y(y)$ | 0 | $m/s$ |
| $y^{bulk}$ | 1 | $mm$ |
| $P^{bulk}$ (isobaric) | 1 | $atm$ |

In addition, a wall BC is required for the mass-fractions, where we have to differentiate between depositing and non-depositing species. Depositing species are subject to a Dirichlet boundary condition by specifying the mass-fraction at the wall;

$$X_{i,w,dep} \approx X_i^{wall} \quad . \tag{16}$$

For non-depositing species, we must require that the convective mass flux is zero at the wall. Neglecting the wall-normal advective velocity, the non-depositing species are thus subject to a Neumann BC specified by

$$j_{i,d,y}\big|_{y^+=0} = 0 \quad . \tag{17}$$

That is, for non-depositing species, the thermophoresis and diffusiophoresis must cancel out, at the wall. The choice of wall BC for the mass-fractions, imply that part of the model formulation is prescribing which species may deposit. Alternatively, a precipitation kinetics model may be employed for the phase-change at the wall.

**SIMULATION RESULTS**

To demonstrate some of the capabilities of the modelling framework, we consider a simplified, semi-artificial, 4-component ideal mixture of perfect gasses, *A,B,C* and *D*, where the *C*-component may sublimate and deposit on the wall. The *D*-component is an inert carrier gas while the three other species obey the reversible chemical reaction

$$2A + B \rightleftharpoons 2C \quad . \tag{18}$$

Material data and considerations about the chemical equilibrium are discussed in the appendix. Flow conditions are given in 2. The model was run for four different scenarios to study the effect of turning on/off deposition and chemical reactions; a) off/off; b) off/on; c) on/off; and 4) on/on.

In Fig. 1, we see that the axial velocity-profile is essentially insensitive to the precipitation/chemical kinetics. As Fig. 2 shows, the temperature profile shows some sensitivity, and in Fig. 3 we see that the fluid mass density strongly affected. The difference between the highest and the lowest mass density values at the wall is approximately 20%. This is partly due to the temperature dependence of the mass-density, but mostly due to the composition dependency. It can be seen that there is a slight difference between the depositing- and non-depositing scenarios, but the main difference is due to the chemical reaction. As is seen in Fig. 4, the mass-fraction profiles are heavily dependent on the modelling premises. In the case of no deposition and no chemical reactions (a), there is a slight variation of the mass-fractions, through the boundary-layer mainly due to the thermophoresis that depends on the molecular weight. Allowing for deposition (c), we see that especially the mass-fraction profile of the depositing species is affected, and we realize that the mass-fraction profiles are strongly dependent on the mass-fraction imposed at the wall, for the depositing species. Allowing for chemical reaction (b,d), we get a huge impact on the mass-fraction profiles. In the case of both deposition and chemical reactions (d) we see that Species A was completely consumed near the wall.

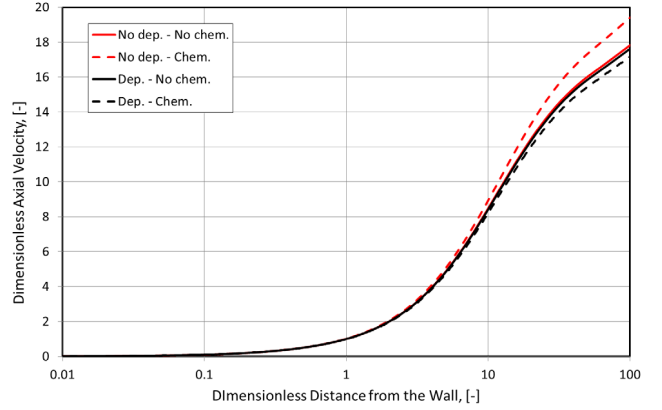

Fig. 1 Test-case dimensionless axial fluid velocity profiles.

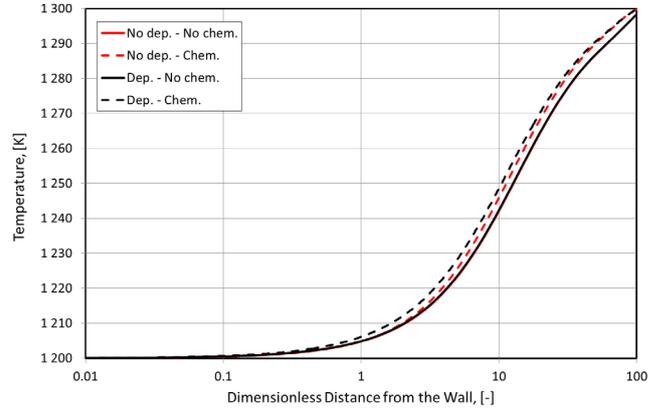

Fig. 2 Test-case fluid temperature profiles.

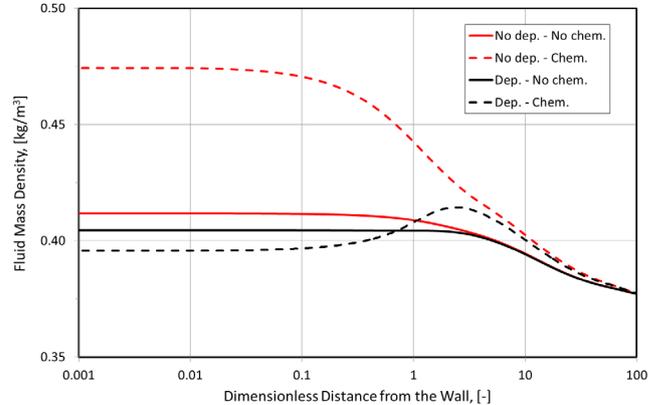

Fig. 3 Test-case fluid mass density profiles.

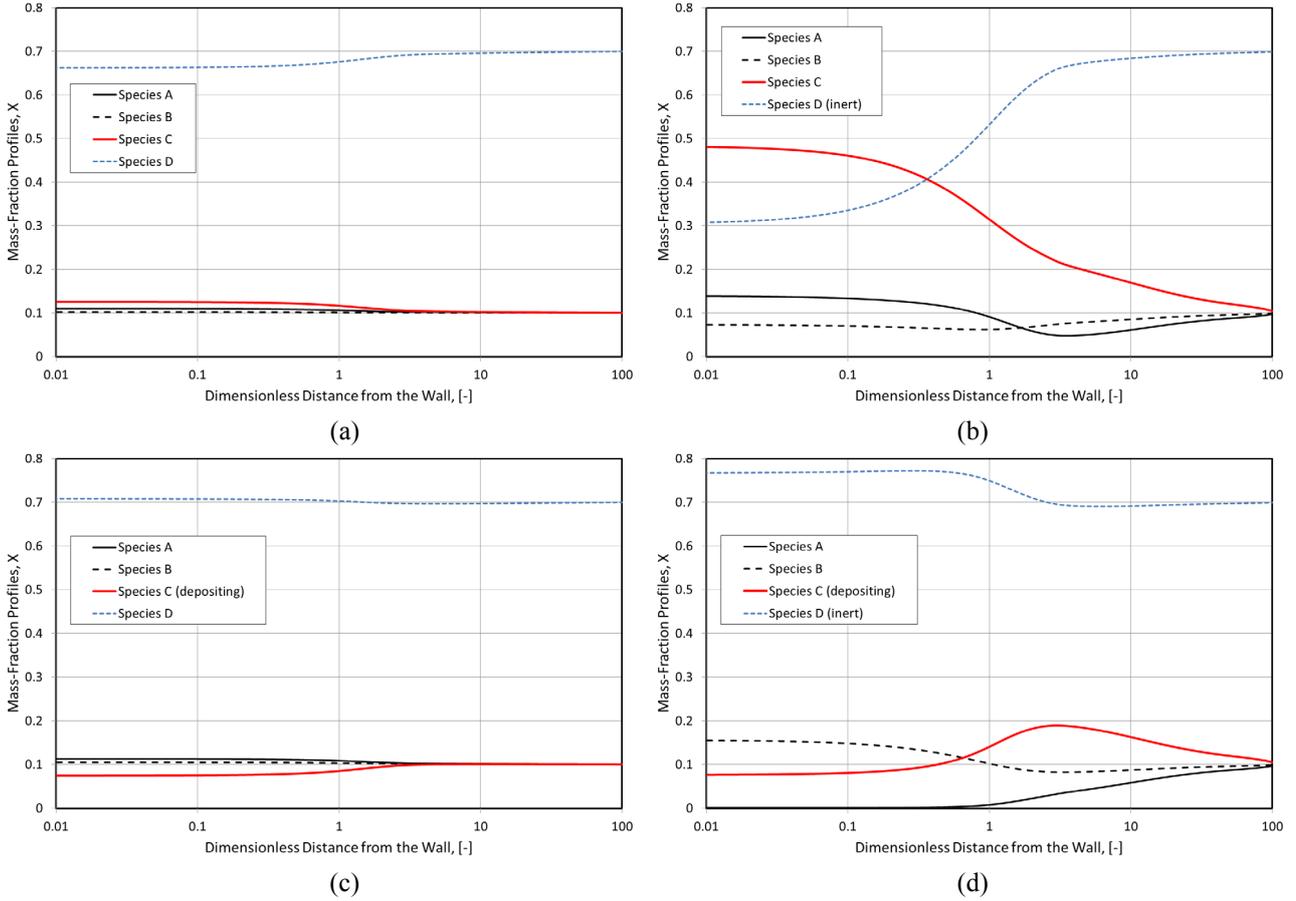

Fig. 4 Test-case mass-fraction profiles for four different scenarios; a) No deposition, and no chemical reactions; b) No deposition, but chemical reaction between species *A,B* and *C*; c) Deposition of species *C*, but no chemical reactions; and 4) Deposition of Species *C*, and chemical reaction between species *A,B* and *C*.

In the case of negligible wall-normal advective velocity, the deposition flux is identical to the diffusion flux of the depositing species at the wall. From Eq.(6) the deposition fluxes of the scenarios c) and d), were found to be $5.8 \cdot 10^{-6}$ and $2.2 \cdot 10^{-3}$ $kg/m^2 s$, respectively.

**DISCUSSION**

The mathematical framework for the modelling of multi-component transport through the turbulent boundary-layer, developed in this paper, is presented in a generic form, and only a few limiting assumptions have been made in the derivation of the governing equations. The main limitations are related to the test-case specific choices of closure models for pure species transport properties.

As a test-case, to demonstrate some of the capabilities of the modelling framework, a simple, semi-artificial system consisting of four species was studied. Although some of the material data were taken from real components, it should be kept in mind that the test-case was set up for demonstrational purposes, only, and not all the physics of the test-case are correct; e.g. the assumption of the chemical reaction being reversible is unlikely, and even the assumption that the product species may exist as a gas component, in the modelled temperature range is dubious, for the imitated system.

**Comments on the Boundary Conditions**

Some of the main assumptions in the model development are related to the boundary conditions. First, it was assumed fully developed flow, and that the far edge of the computational domain was sufficiently far away from the wall that bulk conditions apply. That is, all wall-normal gradients at $y^{bulk}$ are assumed to be zero. If the chosen $y^{bulk}$ is too small, this is not the case, and erroneous solutions may result. Second, the wall BC for the mass-fractions is based on the assumption that the wall-normal advective velocity is negligible. It can be shown that the advective velocity at the wall must be given by $u_{f,y,w} = -j_{i,d,y,w}/(\rho_{f,w} - \rho_{i,w})$, in the case of deposition of one single species, $i$. It follows that the Neumann BC for the non-depositing species, $j$, must be derived from the relation $j_{j,d,y} = -X_{j,w} j_{i,d,y,w}/(1 - X_i^{wall})$. Furthermore, imposing the Dirichlet BC of Eq. (16) on the depositing species is a strong limitation, since it requires us to possess the knowledge of which species will deposit as well as their mass-fractions at the wall. However, the species mass-fractions at the wall are, in reality, functions of temperature, pressure, over-all fluid and foulant compositions, and interphase properties. Thus, the current transport modelling framework requires an additional sub-model for the deposition kinetics at the wall, to obtain correct species mass-fractions at the fluid-foulant interphase.

**Comments on thermodynamics and chemistry**

For the test-case, a simplified thermodynamic description was chosen. Assuming an ideal mixture of perfect gasses is reasonable for gasses at relatively low pressures and high temperatures, but is of course not valid for liquid systems. The assumption of ideal mixing implies that the activity coefficients are identical to one, and this renders the non-ideality tensor, $\Gamma_{jk}$, diagonal. In turn this results in a diagonal matrix product, $D_{ij}\Gamma_{jk}\Lambda_{kl}$, in Eq. (5), reducing the diffusiophoretic term to the Fickian form, where the diffusion flux depends on the species own concentration gradient, only. This is the same result as would be expected in highly diluted mixtures.

Deriving the thermophoretic force as shown in Eq. (5) seems natural from a thermodynamic view (Duhr and Braun, 2006), yet it is not a very popular approach. The reason may be the impracticality in obtaining the pure species absolute entropy, which cannot be derived from thermodynamics alone. The result is that the thermophoretic driving force coefficient must be measured rather than modelled. Sedunov (2001) gives an expression for the absolute entropy of simple liquids/real gasses, but here, risking oversimplifying, and losing generality, we employ the Sackur-Tetrode equation for the entropy of monoatomic ideal gasses (Tetrode, 1912; Sackur, 1913; Grimus, 2013).

In the test-case, a constant value of $\tau_L^+ = 11$ was employed, and throughout the boundary-layer we let $\tau_{char}^+ = \tau_L^+$. That is, the chemical reactions are limited by the turbulent mixing. Close to the wall, however, diffusive mixing may be faster than turbulent mixing, and a more accurate model would be to let $\tau_{char}^+ = \min(\tau_d^+, \tau_L^+)$, indicating that the reaction rates closest to the wall were under-estimated in the test-case. In the case that the chemical reactions cannot be considered mixing limited, a model for the time-scale of the chemical reaction must be implemented, and $\tau_{char}^+ = \min(\tau_d^+, \tau_L^+) + \tau_{chem}^+$ (ANSYS Customer Portal, 2015).

**General comments on the modelling framework**

In all the governing equations, there are advective terms that have been neglected due to the assumption of negligible wall-normal advection. In the case of deposition, the wall-normal velocity is however non-zero and care should be taken to assess this assumption. The low deposition fluxes obtained in the current simulations sustain the assumption.

In the energy equation there is a heat transport term due to the diffusive mass flux density. This term was neglected, without assessing the magnitude of it. Likewise, the enthalpy of fusion due to the fluid-solid transition at the wall was neglected. For small deposition rates, the enthalpy of fusion will have negligible effect on the temperature profile, and this is a reasonable assumption.

Lacking better data, simplified models from kinetic theory was employed to calculate pure component viscosities and thermal conductivities. The Lennard-Jones potential parameters were assumed identical for all the species.

Future work on the modelling framework includes e.g. improvement of the main flaws mentioned above; implementation as a wall boundary-condition sub-grid model in commercially available CFD software; coupling of the current solidification fouling model and previously developed particulate fouling models through thermodynamically driven particle growth/dissolution models; and validation against experimental data.

**CONCLUSIONS**

A generic mathematical modelling framework has been developed and presented for the multi-component mass and heat transport through the turbulent boundary-layer, for fluids that have potential for precipitation fouling and chemical reactions. The model predicts the axial velocity, temperature and mass-fraction profiles, and from these it is possible to calculate the net deposition flux. The modelling framework is well suited for implementation as a sub-grid model acting as a mass sink wall boundary condition in coarse-grid CFD models. Future work will involve implementation of the current modelling framework via user-defined functions, in commercially available CFD software, developing a coupling between the previously developed particulate fouling models and the current direct precipitation fouling model, to handle the growth/dissolution of particles and combined particulate/precipitation fouling.

Some of the capabilities of the current modelling framework were demonstrated by studying a simplified, semi-artificial test-case consisting of an ideal mixture of four perfect gasses capable of chemical reactions and sublimation-deposition on the wall. Four different scenarios were studied with deposition and chemical reactions on/off, and it was shown how the mass-fraction profiles depend on the precipitation/reaction kinetics.

**NOMENCLATURE**

$c_p$  Specific heat capacity, $J/kg\,K$.

$\bar{d}$  Diffusive driving force vector, $1/m$.

$\mathcal{D}$  Binary diffusion coefficient, $m^2/s$.

$D_{ij}$  Diffusivity tensor, $m^2/s$.

$\bar{g}$  Gravity vector, $m/s^2$.

$G_z$  Molar chemical potential, $J/mol$.

$h \equiv q_w/(T^{bulk} - T^{wall})$ Heat transfer coefficient, $W/m^2 K$.

$h_{sens}$  Specific sensible enthalpy, $J/kg$.

$(\Delta h_f)_i^0$ Specific formation enthalpy, $J/kg$.

$\Delta H_r^0$ Standard reaction enthalpy, $J/mol$.

$\bar{j}$  Mass flux density vector, $kg/m^2 s$.

$k_B$  The Boltzmann constant, $1.3807 \cdot 10^{-23}\, J/K$.

$k$  Heat conductivity, $W/mK$.

$K_z^{eq}$ Molar equilibrium constant, *dimensionless*.

$M_w$  Molar mass, $kg/mol$.

$N$  Number of species.

$Nu$ Nusselt number, *dimensionless*.

$P$  Pressure, $Pa$.

$Pr$ Prandtl number, *dimensionless*.
$q_w$ Wall heat flux density, $W/m^2$.
$\mathcal{R}$ The universal gas constant, $8.314\ J/mol\ K$.
$R$ Chemical reaction rate (production), $kg/m^3 s$.
$Re$ Reynolds number, *dimensionless*.
$S$ Molar entropy, $J/mol\ K$.
$Sc$ Schmidt number, *dimensionless*.
$T$ Absolute temperature, $K$.
$u_\tau \equiv \sqrt{\tau_w / \rho_{f,w}}$ Shear velocity, $m/s$.
$\bar{\mathbf{u}}_f$ Mass-averaged advective fluid velocity vector, $m/s$.
$x$ Cartesian coordinate, parallel to the wall, $m$.
$X$ Mass-fraction, $kg/kg$.
$y$ Wall-normal Cartesian coordinate, $m$.
$z$ Mole-fraction, $mol/mol$.
$\gamma$ Activity coefficient, *dimensionless*.
$\Gamma_{jk}$ Non-ideality tensor, *dimensionless*.
$\delta_{ij}$ Kronecker delta, *dimensionless*.
$\partial_\square \equiv \partial/\partial \square$ Differential operator, $1/[\square]$.
$\zeta$ Molar correction to obtain chem. eq., *dimensionless*.
$\Lambda_{kl}$ Mass-mole conversion tensor, *dimensionless*.
$\mu$ Dynamic viscosity, $Pa\ s$.
$\nu$ Kinematic viscosity, $m^2/s$.
$\rho_f$ Fluid mixture mass density, $kg/m^3$.
$\tau_{char}$ Chemical reaction time-scale, $s$.
$\tau_L$ Lagrangian time-scale, $s$.
$\boldsymbol{\tau} = \tau_{\alpha\beta}$ Shear-stress tensor, $Pa$.
$\tau_w$ Wall shear-stress, $Pa$.

**Subscripts**
$b$ Value outside the boundary layer (bulk).
$d$ Diffusive.
$f$ Property of the fluid mixture.
$i, j, k, l$ Species indexing
$t$ Turbulent.
$w$ Value at the wall.

**Superscripts**
$+$ Dimensionless variable.
*bulk* Fixed value outside the boundary layer.
*eq* Chemical equilibrium.
*wall* Fixed value at the wall.

**Averaging notation**
$\bar{}$, $< >$  Mean, ensemble average value.
$\tilde{}$  Favre average.
$'$  Deviation from mean value.
$''$  Deviation from Favre average.

**ACKNOWLEDGEMENTS**
This work was funded by the Research Council of Norway and The Norwegian Ferroalloy Producers Research Association (FFF, 2015), through the SCORE project (Wittgens, 2013). John Morud, Balram Panjwani, Tore Flåtten, and Are Johan Simonsen (all at SINTEF MK) are acknowledged for their invaluable contributions as discussion partners. Sverre especially thanks his family, Maie and Tiril Margrethe, for their patience and support through this hectic period.

# APPENDIX

Fluid mixture properties required in the modelling framework are based on the pure species properties. The model is not limited to any specific way of obtaining the pure species properties, however. These can be calculated from kinetic theory, thermodynamics/chemistry software packages or by interpolating in tables of experimental data.

For the test-case employed in this paper, a combination of calculated properties and tabulated data was employed. Tabulated material data are given in Table 3. The modelled gas system resembles a mixture of SiO (A), $O_2$ (B), $SiO_2$ (C), and Ar (D) in the sense that some of the material properties were taken from these components. Approximate material data were found in (Toropov and Barzakovskii, 1966) and (Chase et al., 1985).

The mass density of the gas mixture was calculated from the ideal gas law;

$$\rho_f = M_{w,f} P / \mathcal{R} T \quad, \tag{19}$$

it was assumed that the gas was calorically perfect, and that the mixture specific heat capacity is calculated as the mass-weighted sum of pure species specific heat capacities,

$$c_{P,f} = \sum_{i=1}^{N} X_i c_{P,i} \quad. \tag{20}$$

The pure component entropies were calculated from the Sackur-Tetrode equation (Tetrode, 1912; Sackur, 1913; Grimus, 2013);

$$S_i(T,P) = \mathcal{R}\left(2.5 \ln T - \ln P + 1.5 \ln M_{w,i} + \sigma_0\right) \quad, \tag{21}$$

where $\sigma_0$ is a constant. The mixture entropy is given by the mole-weighted sum of pure component entropies at their respective partial pressures;

$$S_f(T,P,\{z_i\}) = \sum_{i=1}^{N} z_i S_i(P,T) \quad. \tag{22}$$

The pure-component viscosity and thermal conductivity were estimated from kinetic theory (see e.g. Anderson (2006));

$$\mu_i = 2.6693 \cdot 10^{-6} \sqrt{M_{w,i} T} / d_i^2 \Omega_\mu \quad, \tag{23}$$

$$k_i = \left(c_{P,i} + 1.25 \mathcal{R}/M_{w,i}\right) \mu_i \quad, \tag{24}$$

where we employed identical, constant value Lennard-Jones parameters for all the species, as indicated in Table 3. The mixture viscosity and thermal conductivity are calculated from the pure component properties, by Wilke's rule;

$$\phi_f = \sum_{i=1}^{N} \left(z_i \phi_i / \sum_{j=1}^{N} z_j \Phi_{ij}\right) \quad, \tag{25}$$

where $\Phi_{ij} = \frac{1}{\sqrt{8}} \left(1 + \frac{M_{w,i}}{M_{w,j}}\right)^{-\frac{1}{2}} \left[1 + \left(\frac{\phi_i}{\phi_j}\right)^{\frac{1}{2}} \left(\frac{M_{w,i}}{M_{w,j}}\right)^{\frac{1}{4}}\right]^2$, and

$\phi$ represents either viscosity, $\mu$, or thermal conductivity, $k$. Identical, constant binary diffusion coefficients, $D_{ij} = 1.0 \cdot 10^{-8} \, m^2/s$, were employed for all the species-pairs.

Based on conservation of mole-numbers, it can be shown that the equilibrium mole-fractions are given by

$$z_A^{eq} = \frac{z_A + 2\zeta}{1+\zeta} \quad, z_B^{eq} = \frac{z_B + \zeta}{1+\zeta} \quad, \text{and } z_C^{eq} = \frac{z_C - 2\zeta}{1+\zeta} \quad, \tag{26}$$

where $\zeta$ is found by solving the cubic equation

$$K_z^{eq}(T) = \frac{(z_C - 2\zeta)^2 (1+\zeta)}{(z_A + 2\zeta)^2 (z_B + \zeta)} \quad. \tag{27}$$

It is assumed that the bulk composition (see Table 3) is at chemical equilibrium, so that $K_{z,b}^{eq} = \left(z_C^{bulk}\right)^2 / \left(\left(z_A^{bulk}\right)^2 z_B^{bulk}\right)$, and the temperature-dependent chemical equilibrium constant is given by the van't Hoff equation;

$$K_z^{eq}(T) = K_{z,b}^{eq} \exp\left\{\frac{\Delta H_r^0}{\mathcal{R}}\left(\frac{1}{T_b} - \frac{1}{T}\right)\right\} \quad, \tag{28}$$

where $\Delta H_r^0 = 2 M_{w,C} \left(\Delta h_f\right)_C^0 - 2 M_{w,A} \left(\Delta h_f\right)_A^0 - M_{w,B} \left(\Delta h_f\right)_B^0$ is the standard molar reaction enthalpy of the chemical reaction.

Table 3. Material data for the simulation test-case species.

|  | A | B | C | D |
|---|---|---|---|---|
| $X_i^{bulk}$ | 0.1 | 0.1 | 0.1 | 0.7 |
| $X_i^{wall}$ |  |  | 0.075 |  |
| $M_{w,i}, [kg/mol]$ | 0.044 | 0.032 | 0.06 | 0.04 |
| $c_{P,i}, [J/kgK]$ | 830 | 1125 | 1000 | 520 |
| $d_i, [\text{Å}]$ | 3 | 3 | 3 | 3 |
| $\Omega_i$ | 1 | 1 | 1 | 1 |
| $\left(\Delta h_f\right)_i^0, [J/kg]$ | -2.446E6 | 0 | -5.138E6 |  |